\documentclass{article}
\usepackage{graphicx}
\usepackage{array}
\usepackage{epstopdf}
\usepackage{tabularx}
\usepackage{caption}
\begin{document}
\def\thepage{}

\title{Optimizing weights of protein energy function to improve {\it ab initio} protein structure prediction} 

\author{Chao Wang$^1$, Yi Wei$^1$, Juntao Liu$^{1,3}$, Haicang Zhang$^1$, Bin Ling$^1$, \\
Shuai Cheng Li$^4$, Wei-Mou Zheng$^2$\footnote{Corresponding author. Email: zheng@itp.ac.cn}, Dongbo Bu$^1$\footnote{Corresponding author. Email: dbu@ict.ac.cn}}


\maketitle

\baselineskip=13.5pt

\begin{abstract}

Predicting protein $3$D structure from amino acid sequence remains as a challenge in the field of computational biology.
If protein structure homologues are not found, one has to construct $3$D structural conformations from the very beginning by the so-called {\it ab initio} approach, using some empirical energy functions.
A successful algorithm in this category, Rosetta, creates an ensemble of decoy conformations by assembling selected best short fragments of known protein structures and then recognizes the native state as the highly populated one with a very low energy.
Typically, an energy function is a combination of a variety of terms characterizing different structural features, say hydrophobic interactions, van der Waals force, hydrogen bonding, etc.
It is critical for an energy function to be capable to distinguish native-like conformations from non-native ones and to drive most initial conformations assembled from fragments to a native-like one in a conformation search process.
In this paper we propose a linear programming algorithm to optimize weighting of a total of $14$ energy terms used in Rosetta. 
We reverse  the Monte Carlo  process of Rosetta to approach native-like conformations to a process  generating from the native state an ensemble of initial conformations most relevant to the native state.
Intuitively, an ideal weighting scheme would result in a large ``basin of attraction'' of the native structure, which leads to an objective function for the linear programming.
We have examined the proposal on several benchmark proteins, and the experimental results suggest that the optimized weights enlarge the attraction basin of  the native state and improve the quality of the predicted native states as well.
In addition, a comparison of optimal weighting schema for proteins of different classes indicates that in different protein classes energy terms may have different effects.

\end{abstract}

\baselineskip=13.5pt

\section{Introduction}

Determination of protein structure is important for understanding protein functions\cite{kaiti2}.
The classical techniques for protein structure determination include X-ray crystallography, nuclear magnetic resonance (NMR), and electron microscopy, etc.
These determination techniques, however, suffer from the limitations of both expensive costs and long determination period, leading to the ever-increasing gap between the number of known protein sequences and that of solved protein structures\cite{kaiti7}.
Therefore, computational methods to predict protein structures from sequences are becoming increasing important to narrow down the gap\cite{kaiti3}.\\

Depending on whether protein structure homologues have been empirically solved or not, the protein structure prediction approaches can be categorized into three families: homology modeling\cite{homology,psiblast}, threading\cite{sp3,hhpred,raptor},  and {\it ab initio} methods\cite{rosetta,firstfragment,falcon,tasser,quark}.
Homology modeling approaches exploit the fact that two protein sharing similar sequences often have similar structures, and threading methods compare the target sequence against a set of known protein structures and report the structure with the highest score as the predicted structure.
Although homology modeling and threading approaches generally yield high quality predictions, the two approaches cannot help us understand the thermodynamic mechanism during the the protein folding process.\\

In contrast to homology modeling and threading techniques, {\it ab initio} prediction methods work without requirements of known similar protein structures.
Briefly speaking, {\it ab initio} prediction methods are based on the ``thermodynamic hypothesis'', i.e. the native structure of a protein should be the highly populated one with sufficiently  low energy.
For example, Rosetta\cite{rosetta}, one of the successful {\it ab initio} prediction tools, employs the Monte Carlo strategy to search conformations assembled from fragments of known structures, and finally reports the centroid of a large cluster of low-energy conformations.\\
 
One of the key components of {\it ab initio} prediction approaches is designing an effective energy function\cite{rosetta,falcon,tasser,quark}.
Typically, an energy function consists of a variety of energy terms characterizing different structural features, especially the interplay between local and global interactions among residues.
For example, the hydrophobic interaction term is designed to capture the observed tendency of non-polar residues to aggregate in aqueous solution and exclude water molecules.
Van der Waals force term, is the sum of the attractive or repulsive forces among residues.
Hydrogen bonding term describes the electromagnetic attractive interaction between polar molecules in which hydrogen is bound to highly electronegative atom oxygen in the carboxyl \cite{iupac}.\\

It is critical for an energy function to be able to distinguish native-like conformations from non-native decoy conformations, and drive as much as possible initial  conformations to the native-like one in the conformation search process.
To achieve these two objectives, a widely-used strategy is to maximize the correlation between the energy and the structural similarity of the decoys and the native structure, besides requiring the native structure to be the lowest one.\cite{TOUCHSTONEII}
Inspired by the ``funnel-shaped free energy surface'' idea, Fain et al.\cite{LevittFunnel} proposed a funnel sculpting technique to generate an energy function step by step until a random starting conformation ``roll'' into the native-like neighborhood.
And funnel landscapes describe how protein topology determines folding kinetics. \cite{peter2005}
Shell et al.\cite{shell} attempts to smooth energy function to make the energy landscape a funnel.\\
 
There are usually multiple terms in energy function, e.g. Rosetta utilizes a total of $14$ terms in residue-level conformation search phase, and over $150$ terms in the full-atom mode; therefore, it is important to finding the optimal weighting of so many energy terms.
Schafer et al. \cite{peterreview} proposed a linear programming to ensure the native conformation is more stable than decoys, while in our methods, a series of decoys are sampled to describe the basin near the native conformation.
This study focuses on designing an optimal weighting of a total of $14$ energy terms used in Rosetta.
Central to our effort is the ``reverse sampling'' technique, i.e. Rosetta applies Monte Carlo technique to approach a native-like conformation, while our model attempts to generate from the native state an ensemble of initial conformations most relevant to the native state.
A linear program was proposed to enlarge as much as possible the set of initial conformations that can ``roll'' into the native-like neighborhood.
This way, the possibility of successful search and the quality of predicted conformation increase as the ``basin of attraction'' of the native structure is extended. \\

The manuscript is organized as follows: section $2$ describes the whole framework of our method, and the LP model to optimize protein energy weights as well.
Section $3$ lists experimental results of the optimized energy function. 
In section $4$, we will discuss some limitations of our method and possible future works.
    
\section{Methods}

Our weight optimizing technique works in an iterative manner; that is, we start with a uniform weighting scheme, i.e. all terms have the same weight, and proceed with rounds of alternating estimation of native-like neighborhood and enlargement of the neighborhood.
The two-steps procedure is repeated until the weights change between successive iterations is sufficiently small.
The details of the two steps are described in the following subsections: 

\subsection{Estimating the native-like neighborhood via ``reverse sampling''}
As mentioned above, it is critical for an energy function to drive as much as possible initial assembled conformations to the native-like structures in the conformation search process.
Intuitively, the initial assembled conformations are said to lie in the ``attraction basin'' of the native structure, or native-like neighborhood, if the conformations can finally evolve to the native structure during the search process.
We first describe how to estimate the native-like neighborhood. \\

We utilized a ``reverse sampling'' technique to determine the native-like neighborhood under a specific energy function.
Here, the term ``reverse sampling'' is introduced to describe the conformation generating process essentially reverse to the conformation search process used by {\it ab initio} methods.\\

In particular, {\it ab initio} methods usually employ Monte Carlo technique to search native-like conformations from a random initial conformation.
At each step, a perturbation of the current conformation is made via fragment replacing in Rosetta\cite{rosetta} or torsion angle sampling in FALCON\cite{falcon} to generate a new conformation.
The newly generated conformation is accepted if it has lower energy relative to the original conformation; otherwise the new conformation will be accepted with a probability according to the Metropolis-Hasting rule\cite{montecarlo} (See Figure 1 , left panel  ).
It should be pointed out that starting from some initial conformations, the probability to reach the native-like structures might be  very low. \\

By ``reverse sampling'', we mean that the sampling process starts from the native structure, and attempt to find a neighboring conformation with higher energy at each step (See Figure 1 , right panel  ).
Two constraints were imposed onto the ``reverse sampling'' process: 
\begin{itemize}
\item At each step, the newly generated neighboring conformation should not be too far from the original one. Here, we require that the RMSD (Root Mean Square Deviation) of the two conformations to be less than 0.5 Angstrom. 
\item The reverse sampling process ends at a conformation if we failed to find one of its neighboring  conformations with higher energy after multiple trials, say 1000 times in the study. Informally, this conformation is called ``edge point'' conformation of the attraction basin.
\end{itemize}

This way, a path of conformations $P=S_0 \rightarrow S_1\rightarrow S_2\rightarrow ...\rightarrow S_n$ is generated by the reverse sampling process, where $S_0$ denotes the native structure, $S_i$ $(1\leq i \le n)$ denotes inter-mediate conformations, and $S_n$ denotes the final edge point conformation.
For the edge point conformation $S_n$, its $m$ neighboring conformations are sampled and denoted as $N(S_n)=\{S_n^{(1)}, S_n^{(2)}, ..., S_n^{(m)} \}$. In our study, $m$ is set as 1000. 
The RMSD between $S_0$ and $S_n$ is calculated as a rough measure of the  attraction basin radius. 

\begin{figure}
\centering
\includegraphics[width=3.5in]{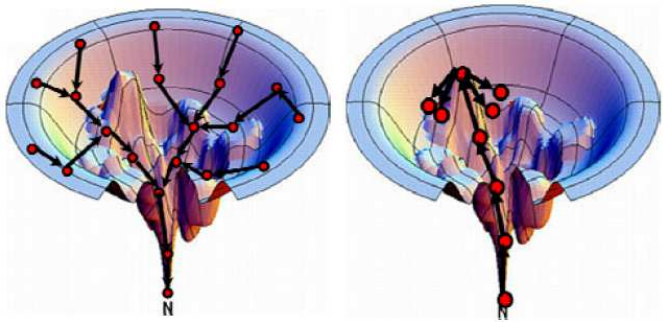}%
\caption{Protein structure prediction usually employs a forward sampling procedure starting from a random conformation to a native-like conformation , left panel  and the reverse sampling process starting from the native structure to estimate the attraction basin , right panel . The reversal sampling of Monte Carlo is to sample a  path of conformations from the native structure. At each step, the energy of decoys increases until reaching an ``edge point'' conformation. A conformation is called ``edge point'' conformation is all of its neighbors has an energy lower than it. The distance between the ``edge point'' conformation and the native structure serves as a measure of the size of the attraction basin. }
\label{ForwardReverseSampling} 
\end{figure}

\subsection{Enlarging attraction basin by cutting off the edge point conformation} 
Intuitively, if we can ``cut off'' the edge point conformation, the attraction basin of the native structure will be enlarged since the reverse sampling process will not stuck at the conformation yet.
The ``cutting off'' operation is accomplished via tuning the weights of the energy terms to decrease the energy of $S_n$ to be less than at least one of its neighbors $S_n^{(i)}$ $(1 \le i \le m)$.
Meanwhile, the new weighting should still keep the order of the inter-mediate conformations in the path, i.e. the energy of $S_i$ should be still lower than that of the conformation $S_{i+1}$. Figure 2 intuitively shows how the conformation path changes during the energy function optimization. 

The weights tuning process is accomplished using the following linear program: 
\begin{eqnarray}
{\rm min}&   &||W-W_0||  \nonumber\\
s.t.\ \ \ \ \ \ \  W \cdot E_i &\leq& W\cdot E_{i+1}, \ \ \ \ \ \ 0\leq i\leq n-1 \\
 W \cdot E_n &\leq& \frac{1}{m}\mathop\sum\limits_{j=1}^{m} W\cdot E_n^{(j)} \\
 W &\geq & 0 \\
 |W| &= & |W_0|
\end{eqnarray}
where the vector $W$ denotes the weights of energy terms, and $W_0$ denotes the original weights.
Thus, the objective of the linear program is to find a new weight with change as small as possible.
For an inter-mediate conformation $S_i$ in the reverse sampling path, $E_i$ denotes the vector of its energy terms, i.e. $E_i= < e_{i}^{(1)},e_{i}^{(2)},\ldots,e_{i}^{(14)} >$.
Formula (1) describes a restriction that the original relative order of $S_i$ and $S_{i+1}$ should be kept even using the new weights, and formula (2) is set to ``cut off'' the edge point conformation, i.e. at least one of the $m$ neighbors of the edge point conformation has a higher energy.
Thus, $S_n$ is no longer an edge point conformation under the new weights. 

\begin{figure}
\centering
\rotatebox{0}{\includegraphics[width=2.5in]{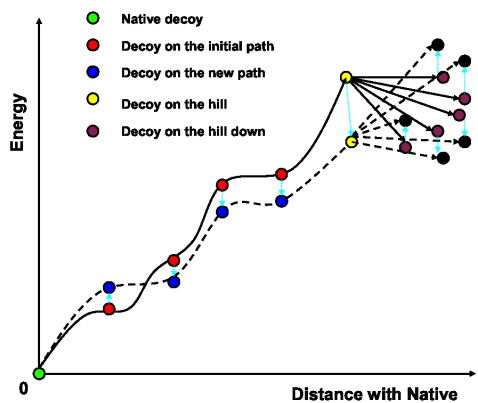}}
\caption{A path of decoys  generated using ``reverse sampling'' technique starting from the native structure. One requirement of the sampling step is that the subsequent conformation should be higher than than the previous one. The reverse sampling step ends at the ``edge point'' conformation (labeled in yellow point), whose neighbors are all lower relatively. Linear program is used to change the weights. The constraints of the linear program are: the order of the conformations in the path should be kept, and at least one neighbors of the ``edge point'' conformation become higher than the edge point conformation itself. This way, the conformation is no longer an ``edge point'' after weight tuning. Intuitively, the attraction basin is enlarge by using the linear program. }
\end{figure}

\section{Results}

\subsection{Data set} 

For proteins in different classes, different energy terms might emphasized.
Thus, we evaluate the weights optimizing procedure on three proteins with different SCOP classes.
The detail information of the three proteins are listed in Table 1. 

\begin{table}
\centering
\label{info}
\begin{tabular}{|c|c|c|c|c|c|}
\hline
PDB ID code & Chain & Class & \#Residues & \#$\alpha$ & \#$\beta$\\
\hline
1ctf & A & $\alpha$+$\beta$ & 68 & 4(38) & 3(18)\\
\hline
1ilo & A & $\alpha$/$\beta$ & 77 & 3(27) & 4(18)\\
\hline
1iie & A & all $\alpha$ & 75 & 3(42) & - \\
\hline
\end{tabular}
\caption{Benchmark protein structures used in the study. The $3$ proteins come from $3$ different SCOP classes: all $\alpha$ (Class A), $\alpha$/$\beta$ (class C), $\alpha$+$\beta$ (class D). Residue numbers are $68$, $77$, $75$, respectively. Columns $5$ and $6$ shows the number and total length of $\alpha$ helices and $\beta$ strands.}
\end{table}

\subsection{Energy terms' weights improve as iteration proceeds} 

In the process of energy function weights optimization, the iteration stops when little changes is observed between consecutive iterations.
Table 2 shows how the weights change as iteration proceeds.
In our experiment, the initial weights of energy terms are set as the weights used by Rosetta in $score3()$.
From this table, it can be observed that after $6$ iterations for protein 1ctfA, the weights are almost fixed.
We can also observed that the difference between consecutive iterations are not very large due to the objective function of the linear program. 
 
\begin{table}
\centering
\begin{tabular}{|c|c|c|c|c|c|c|c|}
\hline
Energy terms & Initial weights & \#1 & \#2 & \#3 & \#4 & \#5 & \#6  \\
\hline
Env     & 1.00 & 1.00 & 0.97 & 0.48 & 0.49 & 1.26 & 1.30 \\
\hline
Pair    & 1.00 & 1.23 & 1.22 & 0.64 & 2.71 & 2.62 & 2.59 \\
\hline
Vdw     & 1.00 & 1.06 & 0.88 & 0.61 & 0.55 & 0.55 & 0.55 \\
\hline
Hs      & 1.00 & 1.70 & 1.70 & 1.06 & 0.09 & 0.09 & 0.09 \\
\hline
Ss      & 1.00 & 1.00 & 1.00 & 0.48 & 0.48 & 0.48 & 0.48 \\
\hline
Sheet   & 1.00 & 1.00 & 1.00 & 1.00 & 3.48 & 3.48 & 3.48 \\
\hline
R-sigma & 1.00 & 1.00 & 1.00 & 0.53 & 0.53 & 0.53 & 0.53 \\
\hline
Cb      & 1.00 & 0.02 & 0.01 & 0.41 & 0.00 & 0.00 & 0.05 \\
\hline
Rg      & 3.00 & 3.00 & 3.14 & 6.05 & 0.49 & 0.49 & 0.49 \\
\hline
Contact Order      & 1.00 & 1.00 & 1.05 & 0.74 & 0.00 & 0.00 & 0.00 \\
\hline
Ramachandran    & 0.00 & 0.00 & 0.00 & 0.01 & 0.91 & 1.01 & 1.02 \\
\hline
Hb-srbb & 0.00 & 0.00 & 0.02 & 0.00 & 3.27 & 2.49 & 2.41 \\
\hline
Hb-lrbb & 0.00 & 0.00 & 0.00 & 0.00 & 0.00 & 0.00 & 0.00 \\
\hline
\end{tabular}
\caption{Initial weights and how the weights change during the iteration process. Here, the initial weights are set as the weights used by Rosetta in $score3()$. A total of 6 iterations are shown here. The Manhattan distances of each adjacent weighting are $1.97$, $0.44$, $6.62$, $17.47$, $1.74$, $0.21$, respectively. The iteration process stops when the Manhattan distance is less than a threshold of $0.3$. The cosine of angle of $\#4$ and $\#5$ weight vector is $0.98$, while that of $\#5$ and $\#6$ reach $0.99$.}
\label{1ctfAIteration}
\end{table}

We also compared the final weights for proteins from different classes (See Table 2). 
Table 2 shows a considerable difference between the optimal weights for proteins from different classes.
For example, environment term in 1iieA (class D) is $5.69$, about $4$ times that in 1ctfA (class D), and over $2.5$ times than that of 1iloA (class C).
This implies that the environment local geometrical term is more important in the all-$\alpha$ class, since only local residue-residue interactions dominate the helix formation.
In addition, 1ctfA (class D) can be distinguished from the other two classes at the ``sheet'' term.
When iteration stops, the ``sheet'' term has a weight of $3.48$ for 1ctfA , much larger than that in 1iloA ($1.90$) and 1iieA ($1.14$).
The main reason is that there are anti-parallel $\beta$-sheets in class D with a $\beta$-hairpin loop inside, instead of $\beta$-$\alpha$-$\beta$ motif in class C.
The table supports the view point that different energy terms are emphasized for proteins in different classes.
Although different proteins adapt different weights of energy function, proteins in the same class obtain similar weights in comparison with those in different classes.\\
	
Table 3 shows the final weights of the above $3$ chains after LP optimization.
After optimization, the weights are quite different from the initial ones.
For example, the weights after iterations  has a Manhattan distance of $15.16$ with the initial ones for protein 1ctfA.\\

The  table also  implies that different energy items play different roles in the protein folding process.
At the very beginning of folding, the hydrophobic interaction dominates the whole process.
Local environment potential leads the whole process for all $\alpha$ proteins.
In comparison, residue pair interaction also plays an important role for $\alpha$+$\beta$ proteins.




\begin{table}
\centering
\begin{tabular}{|c|c|c|c|c|}
\hline
Energy terms   & Initial & 1ctfA & 1iloA & 1iieA \\
\hline
Env     & 1.00 & 1.30 & 1.98 & 5.69 \\
\hline                    
Pair    & 1.00 & 2.59 & 2.02 & 1.09 \\
\hline                          
Vdw     & 1.00 & 0.55 & 0.72 & 0.20 \\
\hline                          
Hs      & 1.00 & 0.09 & 0.82 & 1.28 \\
\hline                          
Ss      & 1.00 & 0.48 & 0.37 & 0.20 \\
\hline                          
Sheet   & 1.00 & 3.48 & 1.90 & 1.14 \\
\hline                          
R-sigma & 1.00 & 0.53 & 1.11 & 0.19 \\
\hline                          
Cb      & 1.00 & 0.05 & 0.24 & 1.62 \\
\hline                          
Rg      & 3.00 & 0.49 & 2.63 & 0.31 \\
\hline                          
Co      & 1.00 & 0.00 & 0.00 & 0.62 \\
\hline                          
Ramachandran    & 0.00 & 1.02 & 0.43 & 0.27 \\
\hline                          
Hb-srbb & 0.00 & 2.41 & 0.80 & 0.36 \\
\hline                          
Hb-lrbb & 0.00 & 0.00 & 0.00 & 0.00 \\
\hline
\end{tabular}
\label{FinalWeights}
\caption{Final weights for protein 1ctfA ($6$ iterations), 1iloA ($5$ iterations), 1iieA ($5$ iterations) are shown in the right $3$ columns. The Manhattan distance of weights between 1ctfA and 1iieA is $15.56$, and that between 1ctfA and 1iloA is $12.39$, while that between 1iloA and 1iieA is only $8.95$. The cosine values of weight vector angle are $0.63$, $0.47$, $0.77$, respectively.}
\end{table}

\subsection{Attraction basin is enlarged during the iteration process}

We also investigated whether the ``cutting off edge point'' strategy helps enlarge the attraction basin of the native structure.
To estimate the attraction basin, a total of 100 paths of conformations were generated using the ``reverse sampling'' technique from the native structure.
The RMSD between the starting native structure and the edge point conformations are calculate as a rough estimation of the radius of the attraction basin.
As shown in Figure 3, the mean RMSD of the $50$ trials is $6$ Angstrom initially, and increases to nearly $14$ Angstrom finally. 
This suggests that the attraction basin is really significantly enlarged during the iteration process. 

\begin{figure}
\centering
\rotatebox{270}{\includegraphics[width=2.5in]{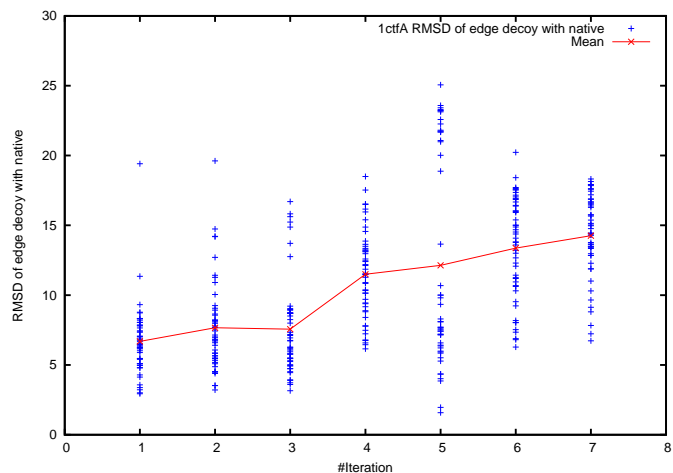}}
\caption{Attraction basin enlarged as iteration proceeds for protein 1ctfA. Here,  $x$-axis denotes the number of iterations, and $y$-axis denotes the RMSD between the native structure and the ``edge point'' conformations. For each weighting scheme, a total of 100 paths were generated, and there are 100 points at each iteration step. The mean RMSD increases as iteration proceeds which implies that the ``cutting off edge point'' strategy really help enlarge the attraction basin. }
\label{EnlargeBasin} 
\end{figure}

\subsection{Optimized energy function help improve protein structure prediction} 

Finally we conducted experiments to investigate whether the optimized weights help improve protein structure prediction or not.
To achieve this goal, we run Rosetta for the testing proteins with weights obtained at each iteration step.
For each weighting scheme, a total of 1000 decoys were generated by Rosetta.
Among these decoys, a clustering procedure is run and the centroid of the largest cluster is reported as the final prediction.
The ``good decoy ratio'' is also calculated.
Here, we adopted a widely-used criteria that a decoy is called ``good decoy'' if it has a RMSD less than $6$ Angstrom to the native structure\cite{falcon}.\\

Taking protein 1iloA as an example, the best prediction has a RMSD of $2.7$ Angstrom (See Figure 6, left panel ), and as iteration proceeds the quality of the best prediction improves step by step and finally the best prediction has a small RMSD of only $1.3$ Angstrom (See Figure 6 , right panel ).
Similar trends were observed for protein 1ctfA and 1iloA (See Figure 4). \\

In addition, the ``good decoy ratio'' was also improved significantly (See Figure 5).
For example, if using the initial weights, there are only 10 good decoy among the $1000$ decoys generated by Rosetta for protein 1iloA; in contrast, there are over $200$ good decoys among the $1000$ generated decoys when using  the optimized weights.
This means that using the optimized weights of energy function, Rosetta can generate high-quality decoys more efficiently. \\

\begin{figure}
\centering
\rotatebox{270}{\includegraphics[width=2.5in]{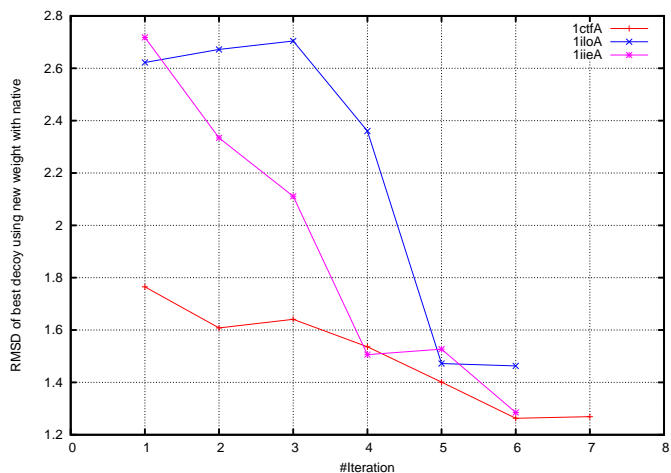}}
\caption{The quality of best prediction increases as iteration proceeds. At each iteration step, a total of 1000 decoys were generated by Rosetta with corresponding weights. We run clustering procedure for the $1000$ decoys and finally select the centroid of the largest cluster as the best decoy. 
The RMSD of the best decoys are calculated,  demonstrating that the weights of energy terms become better and better as iteration proceeds. }
\label{BestDecoy}
\end{figure}

\begin{figure}
\centering
\rotatebox{270}{\includegraphics[width=2.5in]{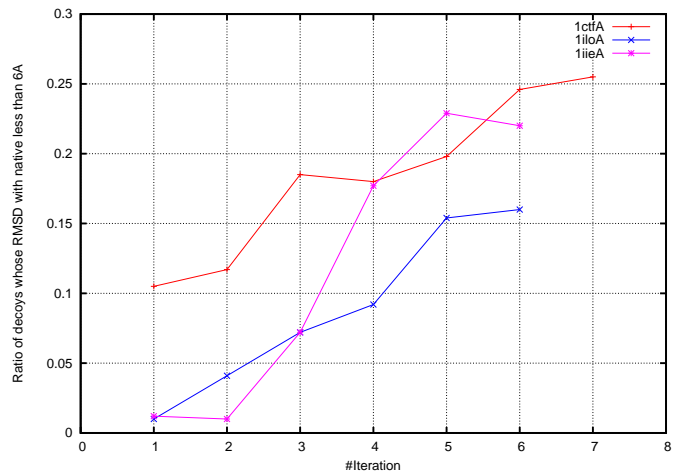}}
\caption{``Good decoy ratio'' increases as iteration proceeds. At each iteration step, a total of $1000$ decoys were generated by Rosetta using corresponding weights of energy terms. Here a decoy is called ``good decoy'' if it has a RMSD less than $6$ Angstrom to the native structure. The figure suggests that the ``good decoy ratio'' significantly increases, e.g. the ratio increases from $0.01$ to over $0.2$ for protein 1iloA. Thus, Rosetta can generate high-quality decoys more efficiently. }
\label{GoodDecoy}
\end{figure}

 \begin{figure}%
   \begin{center}%
     \begin{minipage}{0.25\textwidth}%
      \includegraphics[width=1.0\textwidth]{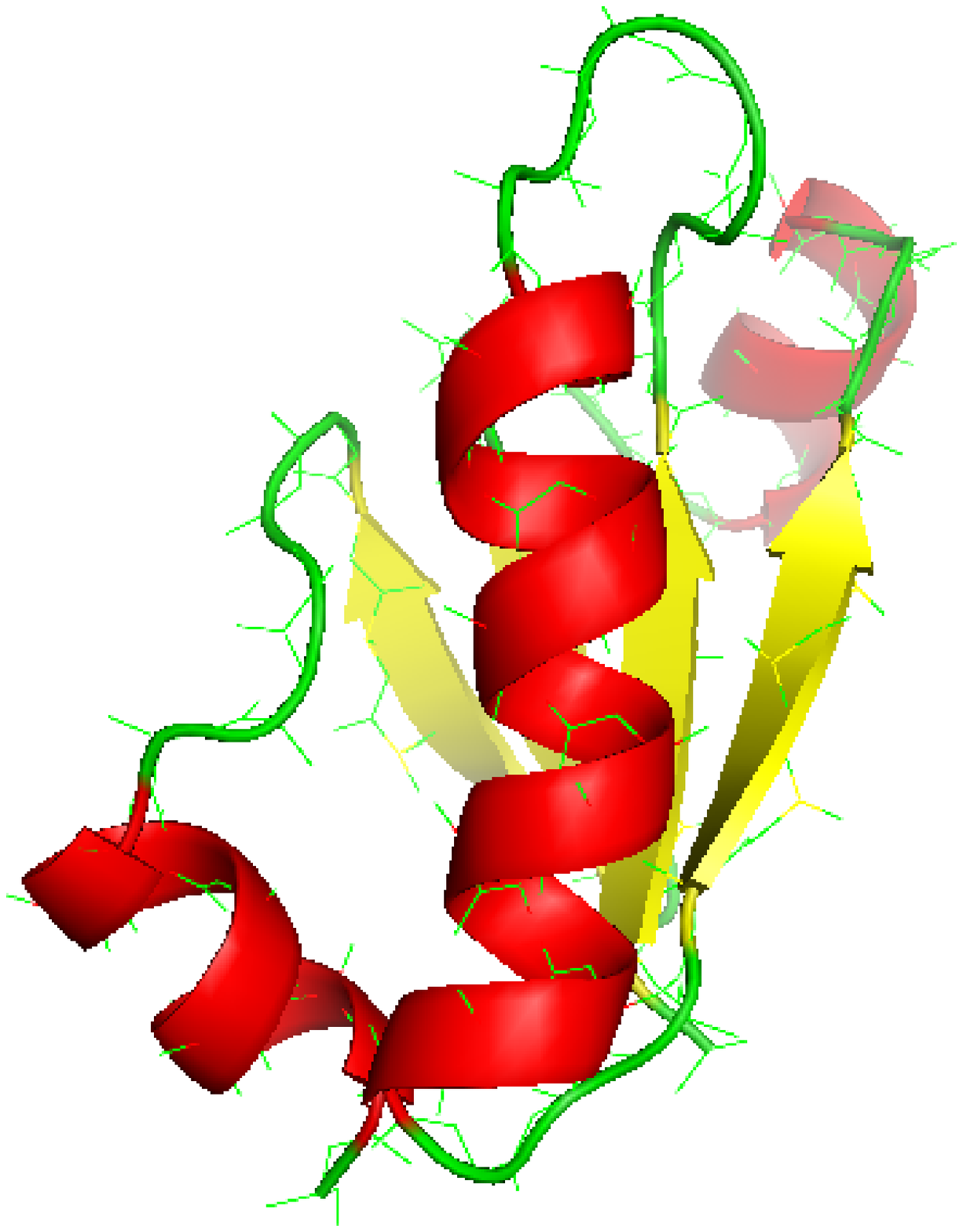}%
     \end{minipage}%
     \begin{minipage}{0.32\textwidth}
      \includegraphics[width=1.0\textwidth]{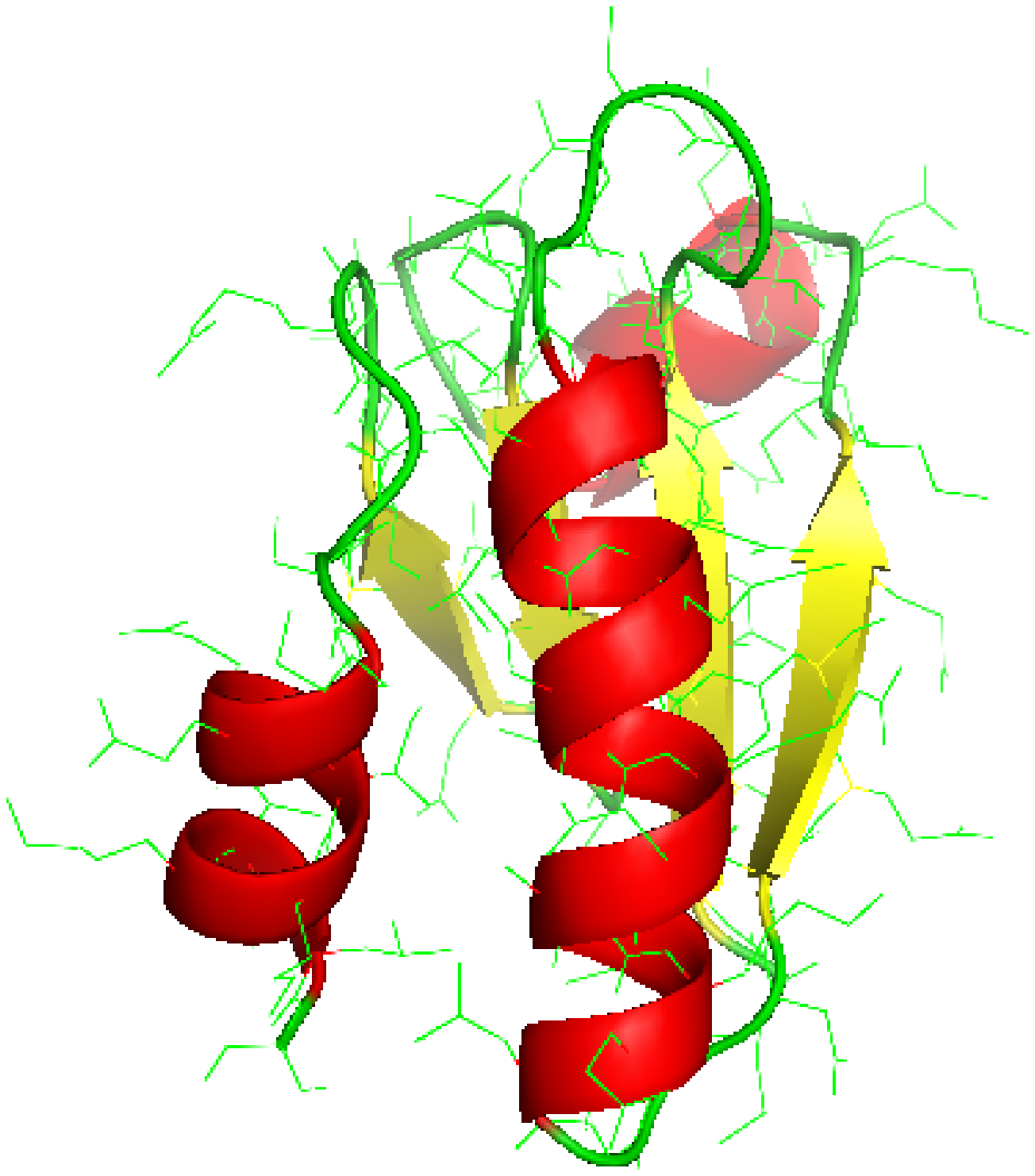}
      \end{minipage}%
     \begin{minipage}{0.25\textwidth}%
      \includegraphics[width=1.0\textwidth]{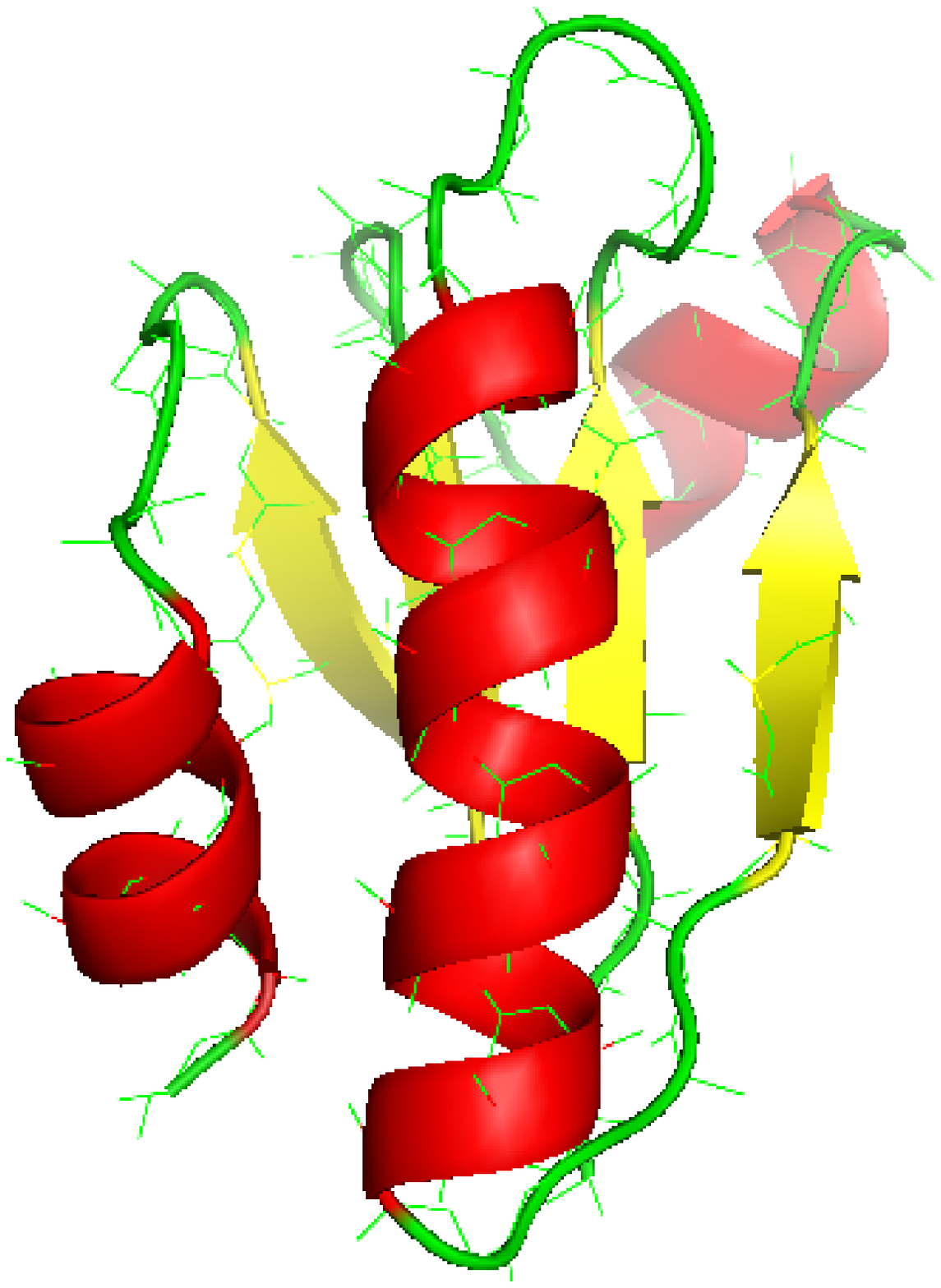}%
     \end{minipage}%
   \end{center}
   \label{1iloA}
   \caption{Native structure (middle panel) of protein 1iloA. The predicted structures using the initial weight , left panel  and the optimal weights , right panel  are also shown. The RMSD are $2.7$ Angstrom and $1.3$ Angstrom, respectively. Thus, the optimized weights help improve the quality of predicted structures.}
 \end{figure}

%
%
%

\section{Conclusion and discussion}

In the study, we present an attempt to find the optimal weights of energy terms.
The basic idea is to estimate the attraction basin of the native structure using ``reverse sampling'' technique, and then enlarge the attraction basin using a linear program.
Experimental results on several benchmark proteins suggest that the optimized weights can significantly improve Rosetta's prediction, and the prediction efficiency is substantially increased as well. \\

It has been reported that the energy terms apply at different stages of protein folding.
According to this observation, Rosetta employs a multi-step prediction strategy.
In particular, Rosetta first uses score function $score_0$ with only hydrophobic core terms, then uses $score_2/score_5$ with secondary structure terms, and finally uses $score_3$ to incorporate more energy terms.
The study here focuses on the optimization of weights for the third step. \\

Our work focuses on the attraction basin near the native structure.
When we get a conformation with over $12$ Angstrom to the native structure, the reverse sampling ends since a conformation with so large RMSD is usually a random conformation.
The random conformations are excluded since they provide little information for the attraction basin. \\

One of the limitations of the study presented here is the limited number of benchmark proteins.
The weights were trained for each protein individually.
Ideally, we have only a common weighting scheme for a protein class rather than a specific weighting for an individual protein.
How to extend the linear program to achieve this objective remains as one of our future works. \\



\section*{Acknowledgement}
This study was funded by the National Basic Research Program of China (973 Program 2012CB316502), the National Natural Science Foundation of China (30800189, 30800168, 90914047), and the Strategic Priority Research Program of the Chinese Academy of Sciences, Grant No. XDA06030200. WM Zheng was supported in the part by National Natural Science Foundation of China (11175224, 11121403).

\end{document}